
\magnification=1200
\font\titlea=cmb10 scaled\magstep1

\baselineskip=12pt
\rightline{IC/93/13}
\rightline{UG-1/93}
\rightline{hepth@xxx/9301085}
\baselineskip=18pt
\vskip .5cm
\centerline{\titlea On the Symmetry Algebra of the Discrete States}
\centerline{\titlea in $d < 2$ Closed String Theory}
\vskip 2cm
\baselineskip=14pt
\centerline{Sudhakar Panda}
\bigskip
\centerline{\it Institute for Theoretical Physics}
\centerline{\it University of Groningen}
\centerline{\it Nijenborgh 4, 9747 AG Groningen}
\centerline{\it The Netherlands}
\bigskip
\centerline{and}
\bigskip
\centerline{Shibaji Roy}
\bigskip
\centerline{\it International Centre for Theoretical Physics}
\centerline{\it Trieste - Italy}
\vskip 1.0cm
\baselineskip=18pt
\centerline{\titlea Abstract}
\bigskip
The symmetry charges associated with the Lian-Zuckerman states for $d < 2$
closed string theory are constructed. Unlike in the open string case, it is
shown here that the symmetry charges commute among themselves and act
trivially on all the physical states.
\vfill
\eject
\noindent{\bf I. Introduction}

In a beautiful paper, Lian and Zuckerman studied the basic building blocks
of the physical states or the BRST cohomology for the $(p,q)$ minimal models
coupled to two dimensional gravity [1]. Their stimulating result showed that
apart from the usual physical states at ghost number zero [2,3]
(with the convention
that the $SL(2,C)$ invariant vacuum has ghost number $-1$), there are infinite
number of physical states at different ghost numbers corresponding to the
null vectors present in each conformal block. Some of these physical states
at low ghost numbers have been constructed explicitly. For higher ghost
number states, the explicit construction is more involved since one has to
deal with null vectors at higher levels. Among the low lying ghost number
states, the form of the physical states (Lian-Zuckerman states) at ghost
number ($-1$) (or physical operators at ghost number zero) are given in [4,5].
The short distance behaviour of the operator product expansions of these
physical operators defines an interesting ring structure, the so called
`ground ring'[6]. The importance of this algebraic structure, in the context
of $d=2$ string theory has been emphasized in [7,8]. In ref. [9] it is observed
that for $d <2$ string theory, there exists another generator $w$ with ghost
number ($-1$) and its inverse $w^{-1}$ with ghost number one. Furthermore, they
proved that any power of $w$ gives a non-trivial cohomology and thus the
ghost number $n$ sector is described by $R_n~=~w^{-n}~R_0$ where $R_0$ is the
ring of ghost number zero physical operators. The consequence of this is that
for minimal models coupled to gravity, the full ring structure is $R~=~R_0
\otimes C [w,w^{-1}]$. This is found to be non-commutative since the generator
$w$ anticommutes with the generators of $R_0$, which are usually denoted by
$x$ and $y$. In a different approach, the physical states with non-zero ghost
numbers are shown to be related to the ghost number zero states (Dotsenko-
Kitazawa states [10]) by making use of the
descent equations of double-cohomology
(i.e. the usual string BRST cohomology and Felder's BRST cohomology) [11].
However, these latter type of physical states do not lie inside the primary
conformal grid.

In the context of $d =2$ string theory, it has been pointed out by Witten [7]
that some special physical operators which belong to the zero eigenvalue space
of the operator $b_0$ give rise to the conserved currents modulo BRST
commutator. In fact, it is shown that the conserved charges corresponding to
the ghost number (1,0) and (0,1) operators generate an area preserving
diffeomorphism in the $x-y$ plane. This result is the reminiscent of the
$W_{\infty}$ symmetry one encounters in $c = 1$ matrix models [12--15], but the
clear relationship between these two $W_{\infty}$ symmetries is still lacking.
For $(p,q)$ minimal models coupled to gravity, as we have mentioned the
picture is quite different. However, it is probably reasonable to expect a
$W^{(q)}$-type symmetry among the conserved charges in analogy with the
$(q-1)$ matrix model.

In the $d <2$ open string case, the vector fields associated with the physical
states have been constructed in ref.[9] and it is shown that their algebra
contains a Virasoro algebra as the subalgebra. In this paper, we construct
the vector fields for $d <2$ closed string physical states, at arbitrary ghost
numbers, belonging to both relative (annihilated by $b_0$) and semi-relative
(annihilated by $b_0 - \bar b_0$) cohomology. We show that the vector fields
commute among themselves unlike in the open string case. This in turn implies
that the conserved currents act trivially on all the physical states. Hence it
seems there is no new symmetry associated with these discrete states in the
free closed string in $d < 2$.

The organization of our paper is as follows. In sec.II, we briefly review
the physical state spectrum for $d < 2$ string theory and fix our notations
and conventions. The vector fields of the conserved charges associated
with the physical states are constructed in sec.III. In sec.IV, we present our
conclusions.

\noindent{\bf II. Physical State Spectrum for $d < 2$ Closed String Theory:}

The $(p,q)$ minimal models (where $p > q$ and $gcd (p,q) = 1$) coupled to
two dimensional gravity can be described in terms of Coulomb gas representation
where the energy momentum tensors for the matter and Liouville sector are
given as
$$
\eqalign{T_M(z)~=&-{1\over 2}~:~\partial X~\partial X~:~+~i Q_M \partial^2 X\cr
         T_L(z)~=&-{1\over 2}~:~\partial \phi~\partial \phi ~:~+~i Q_L
\partial^2 \phi\cr}
\eqno(2.1)
$$
Here $X$ and $\phi$ represent the matter and Liouville fields respectively,
whereas $2Q_M$ and $2Q_L$ denote the background charges. Since, we are working
in the free theory with zero cosmological constant, we can concentrate only
on the chiral sector and then finally combine the left and right moving part
in order to obtain the full closed string spectrum. The Virasoro central
charges for the matter and Liouville part have the form
$$
\eqalignno{&c_M~=~1~-~12 Q_M^2 &(2.2a)\cr
           &c_L~=~1~-~12 Q_L^2 &(2.2b)\cr}
$$
The vertex operators $e^{i p_M X}$ and $e^{i p_L \phi}$ have conformal
weights $\Delta (p_M) =~{1\over 2} p_M (p_M - 2Q_M)$ and $\Delta (p_L) =~
{1\over 2} p_L (p_L -2Q_L)$ respectively. The screening charges for the
matter and Liouville sectors are given by
$$
\eqalignno{&\alpha_{\pm}~\equiv p_M^{\pm}~=~Q_M~\pm~\sqrt{Q_M^2 + 2} &(2.3a)\cr
           &\beta_{\pm}~\equiv p_L^{\pm}~=~Q_L~\pm~\sqrt{Q_L^2 + 2} &(2.3b)\cr}
$$
{}From the above equations, we have $\alpha_++ \alpha_-~= 2 Q_M \equiv
\alpha_0$;
$\alpha_+ \alpha_-~=-2$ and similarly $\beta_+ +\beta_-~= 2 Q_L \equiv \beta_0$
and $\beta_+ \beta_-~= - 2$. Also, since the matter sector is the $(p,q)$
minimal models which are characterized by the Virasoro central charge
$ 1 - {6(p-q)^2\over pq}$, it, therefore, follows from (2.2a) that $2 Q_M =
{}~\sqrt{{2p\over q}} - \sqrt{{2q\over p}}$ and the screening charges take the
values
$$
\eqalign{\alpha_+&=~\sqrt{{2p\over q}}\cr
         \alpha_-&=~- \sqrt{{2q\over p}}\cr}
\eqno(2.4)
$$
Since the central charge of the combined matter and Liouville system
should add up to 26, we also have the relation from (2.2) that
$$
Q_M^2~+~Q_L^2~=~- 2
\eqno(2.5)
$$
{}From this, in terms of $p, q$, the background charge for the Liouville sector
is given by $2 Q_L =~i (\sqrt{{2p\over q}}~+~\sqrt{{2q\over p}})~=~i (\alpha_+
-\alpha_-)\equiv (\beta_++ \beta_-)~=~\beta_0$. Therefore, the screening
charges for the Liouville sector take the values
$$
\eqalign{\beta_+&=~i \alpha_+~=~i \sqrt{{2p\over q}}\cr
         \beta_-&=~-i \alpha_-~=~i \sqrt{{2q\over p}}\cr}
\eqno(2.6)
$$
Let us also recall that the primary fields of the $(p,q)$ minimal models
are represented by the vertex operators $e^{i\alpha_{m,m'}X}$ having the
conformal weights
$$
\Delta (\alpha_{m,m'})~=~{1\over 2} \alpha_{m,m'} (\alpha_{m,m'} -2 Q_M)
{}~=~{1\over 4pq} [(pm - qm')^2 - (p - q)^2]
\eqno(2.7)
$$
where $1\leq m \leq q-1,~~1\leq m' \leq p-1$ and choosing the negative root
for $\alpha_{m,m'}$ we have
$$
\alpha_{m,m'}~=~{1\over 2} \bigl [ (1-m)\alpha_+~+~(1-m')\alpha_-\bigr ]
\eqno(2.8)
$$
The irreducible highest weight module of the matter sector is obtained by
quotienting the Verma module over each primary by its maximum proper
submodule. The submodule is generated by a pair of null vectors
associated with the primary field. The irreducible Virasoro module can be
represented by the following embedding diagram, $E_{m,m'}$:
$$
\def\sp{\nearrow\!\!\!\!\!\!\searrow}
\matrix{&&a_{-1}&\longrightarrow&e_{-1}&\longrightarrow&a_{-2}
&\longrightarrow&e_{-2}&\longrightarrow&\cdots\cr
e_0&\nearrow\atop\searrow&&\sp&&\sp&&\sp&&\sp&\cdots\cr
&&a_0&\longrightarrow&e_1&\longrightarrow&a_1&\longrightarrow&e_2
&\longrightarrow&\cdots\cr}
$$
where
$$
\eqalign{a_t&=~{1\over 4pq}~[ (2pqt + pm + qm')^2~-~(p -q)^2 ]\cr
         e_t&=~{1\over 4pq}~[ (2pqt + pm - qm')^2~-~(p -q)^2 ]\cr}
\eqno(2.9)
$$
Each node of the above diagram represents a Verma module with $a_t$ or
$e_t$ as the dimension of its highest weight state and are null vectors
over $e_0$. An arrow connecting two spaces $E\to F$ means that the module
$F$ is contained in the module $E$. It was proved by Lian and Zuckerman [1]
and also by Bouwknegt, McCarthy and Pilch [5] that there exists a unique
physical state of the combined matter, Liouville and ghost system at each
of these values of $a_t$ and $e_t$ if and only if the Liouville momenta
satisfy the relation:
$$
\Delta (p_L)~=~\cases{1~-& $a_t$\cr
                      1~-& $e_t$\cr}
\eqno(2.10)
$$
and the ghost number of the state is given as
$$
n_{gh}~=~\pi (p_L)~d (p_L)
\eqno(2.11)
$$
where $\pi (p_L)~=~sign~[ i (p_L - Q_L) ]$ and $d (p_L)$ is the number of
arrows from the top node $e_0\equiv \Delta (m,m')$ to that particular node
$a_t$ or $e_t$. This proof is based on the BRST quantization scheme where
the states are taken to be in the space of conformal fields such that a
physical state is BRST closed and it belongs to the zero eigenvalue space
of $L_0^{tot}$. The BRST charge is defined as
$$
Q~=~\sum_{n\in Z} c_{-n} L_n^{(L+M)} ~-~{1\over 2} \sum_{m,n\in Z}~(m - n)~
: c_{-m} c_{-n} b_{m+n} :
\eqno(2.12)
$$
where $b,~c$ are the usual reparametrization antighost and ghost with conformal
weight 2 and $- 1$; and ghost number $- 1$ and 1 respectively. The Virasoro
generators are written in terms of the oscillators as
$$
\eqalignno{L_n^{(L)}&=~{1\over 2} \sum_{m\in Z} :\phi_m \phi_{n-m} :~- (n+1)
Q_L \phi_n&(2.13a)\cr L_n^{(M)}&=~{1\over 2} \sum_{m\in Z} :\alpha_m \alpha_{
n-m} :~-(n+1) Q_M \alpha_n&(2.13b)\cr L_0^{tot}&=~L_0^{(L+M)}~+
\sum_{m\in Z}~m~: c_{-m} b_m :&(2.14)\cr}
$$
In the space of conformal fields, a general physical operator takes the form
$$
{\cal O}~=~{\cal P} (\partial X, \partial \phi, b, c )~e^{i\alpha_{m,m'} X}
{}~e^{i\beta_{n,n'} \phi}
\eqno(2.15)
$$
where, because of the zero eigenvalue of $L_0^{tot}$, ${\cal P}$ is a
differential polynomial of conformal weight
$-~{1\over 2} \alpha_{m.m'} (\alpha_{m,m'} - 2 Q_M)~-~{1\over 2} \beta_{n,n'}
(\beta_{n,n'} - 2 Q_L)$. Here $\alpha_{m,m'}$ is as given in (2.8) and
$$
\beta_{n,n'}~=~{1\over 2} [ (1 - n) \beta_+~+~(1 - n') \beta_- ]
\eqno(2.16)
$$
are the Liouville momenta which are the solutions of (2.10). We note that
$\alpha_{m,m'}$ has the symmetry $\alpha_{m\pm q, m'\pm p} = \alpha_{m,m'}$
whereas, $\beta_{n,n'}$ has different symmetry property namely,
$\beta_{n\pm q, n'\mp p} = \beta_{n,n'}$. Also we have $\alpha_{m,m'} +
\alpha_{n,n'} = \alpha_{m+n-1, m'+n'-1}$ and $\beta_{m,m'} + \beta_{n,n'}
= \beta_{m+n-1, m'+n'-1}$.

The allowed Liouville momenta of the physical operators in (2.15), obtained
by solving (2.10) can be summarized as follows:
$$
\eqalign{p_L^\pm (a_{-t})&=~\beta_{\mp qt \pm m, \mp pt \pm m'}\cr
        p_L^\pm (a_{t-1})&=~\beta_{\mp q(t-1) \mp m, \mp p(t-1) \mp m'}\cr
p_L^\pm (e_t)&=~\beta_{\mp qt \mp m, \mp pt \pm m'}\cr
p_L^\pm (e_{-t})&=~\beta_{\mp qt \pm m, \mp pt \pm m'}\cr
p_L^\pm (e_0)&=~\beta_{\mp m, \pm m'}\cr}
\eqno(2.17)
$$
where $t > 0$, and $p_L^\pm$ refer to the Liouville momentum greater than
$Q_L$ or less than $Q_L$. The presence of the generator $w$ can be seen by
noting that all the allowed Liouville momenta can be split into three
parts, each part associated with one generator, as follows:
$$
\eqalignno{p_L^+ (a_{-t}) &=~- 2t \beta_w~+~(m'-1) \beta_x~+~(m-1) \beta_y
&(2.18a)\cr p_L^+ (a_{t-1}) &=~- 2t \beta_w~+~(p-m'-1) \beta_x~+~(q-m-1)
\beta_y &(2.18b)\cr p_L^+ (e_{-t}) &=~- (2t+1) \beta_w~+~(p-m'-1) \beta_x~+~
(m-1) \beta_y &(2.18c)\cr p_L^+ (e_t) &=~- (2t+1) \beta_w~+~(m'-1) \beta_x
{}~+~(q-m-1) \beta_y &(2.18d)\cr p_L^+ (e_0) &=~- \beta_w~+~(m'-1) \beta_x~+~
(q-m-1) \beta_y &(2.18e)\cr p_L^- (a_{-t}) &=~(2t-2) \beta_w~+~(p-m'-1) \beta_x
{}~+~(q-m-1) \beta_y &(2.18f)\cr p_L^- (a_{t-1}) &=~(2t-2) \beta_w~+~(m'-1)
\beta_x~+~(m-1) \beta_y &(2.18g)\cr p_L^- (e_{-t}) &=~(2t-1) \beta_w~+~(m'-1)
\beta_x~+~(q-m-1) \beta_y &(2.18h)\cr p_L^- (e_t) &=~(2t-1) \beta_w~+~(p-m'-1)
\beta_x~+~(m-1) \beta_y &(2.18i)\cr p_L^- (e_0) &=~- \beta_w~+~(p-m'-1) \beta_x
{}~+~(m-1) \beta_y &(2.18j)\cr}
$$
where again we have taken $t > 0$ and we have defined $\beta_x = \beta_{1,2}
= - {1\over 2} \beta_- = - {i\over 2} \sqrt{{2q\over p}}$; $\beta_y =
\beta_{2,1} = - {1\over 2} \beta_+ = - {i\over 2} \sqrt{{2p\over q}}$ and
$\beta_w = \beta_{q+1,1} = \beta_{1,p+1} = -{i\over 2} \sqrt{2pq}$. Few
comments are in order now. First of all, we note that as we fix the matter
sector $(m,m')$ with $1\leq m\leq q-1$ and $1\leq m'\leq p-1$ in (2.15), the
Liouville sector $(n,n')$ is no longer arbitrary but is getting fixed by
(2.10). Secondly, the coefficient of $\beta_w$ in (2.18) is precisely the
ghost number of the operators appearing at different nodes ($a_t$ or $e_t$)
of the embedding diagram as can be obtained from (2.11). Finally, it is clear
from (2.18) that the whole spectrum of physical states is generated by the
operators $x$, $y$ and $w$ (or $w^{-1}$). Since $\beta_x$ and $\beta_y$ are not
associated with the ghost number, the physical operators $x$ and $y$ are
spin zero, ghost number zero operators and are the generators of the ground
ring $R_0$. In our conventions, their explicit forms are
$$
\eqalignno{x&=~[ bc~+~{3\over 4} \sqrt{{2q\over p}}~(i\partial X + \partial
\phi) ]~e^{i\alpha_{1,2} X~+~i\beta_{1,2} \phi} &(2.19)\cr y&=~[ bc~-~
{3\over 4}\sqrt{{2p\over q}}~(i\partial X - \partial \phi) ]~e^{i\alpha_{2,1}
X~+~i\beta_{2,1} \phi} &(2.20)\cr}
$$
The other generators $w,~w^{-1}$ in general would have the form,
$$\eqalign{w&=~{\cal P} (\partial X, \partial \phi, b, c)~e^{i \alpha_{q-1,1}
X~+~i \beta_{1, p+1} \phi}\cr w^{-1}&=~c e^{i \alpha_{q-1,1} X~+~i \beta_{1,
p+1} \phi}\cr}
\eqno(2.21a)
$$
or,
$$
\eqalign{w&=~{\cal P} (\partial X, \partial \phi, b, c)~e^{i \alpha_{1,p-1} X
{}~+~i \beta_{q+1,1} \phi}\cr w^{-1}&=~c e^{i \alpha_{1,p-1} X~+~i
\beta_{1,-p+1}
\phi}\cr}
\eqno(2.21b)
$$
where ${\cal P}$ is a differential polynomial of conformal weight $p + q - 1$
and ghost number ($-1$). It is clear from $(2.18g)$ that the operators $x$ and
$y$ belong to the matter sector (1,2) and (2,1) respectively and appear at the
position of $a_0$, for $p_L < Q_L$, in the embedding diagram. Also from
$(2.18h)$ and $(2.18i)$, we find that $w$ belongs to either in $(q-1,1)$ or
in $(1,p-1)$ matter sector as given in $(2.21a)$ and $(2.21b)$ and appears at
$e_{-1}$ or $e_1$, accordingly, for $p_L < Q_L$. Similarly, from $(2.18j)$
and $(2.18e)$, we see that $w^{-1}$ belongs to the matter sector $(1,p-1)$ or
$(q-1,1)$ for $p_L < Q_L$ or $p_L > Q_L$ respectively and appears at the node
$e_0$ of the embedding diagram. We note that since $w\cdot w^{-1}~\sim~$I, so
their multiplication is well defined if we take $w$ from $(2.21a)$ and
$w^{-1}$ from $(2.21b)$ or vice-versa. Notice from $(2.18 g,f)$ and
$(2.18 a,b)$ that all even powers of $w$ and $w^{-1}$ belong to the matter
identity sector. Similarly from the other equations in (2.18), it follows
that odd powers of $w$ and $w^{-1}$ belong to either $(1,p-1)$ or $(q-1,1)$
matter sector. Since the explicit form of $w$ will involve constructing the
operator ${\cal P}$ at level $(p+q-1)$, so in general it will be quite
complicated. But for small values of $p$ and $q$, it can indeed be
constructed; for example, for (3,2) model (pure Liouville gravity), it has
the form [16,17,9]
$$
w~=~(\partial^2 b - 3 \partial b b c - {\sqrt{3}\over 2} \partial b\partial
\phi + {\sqrt{3}\over 2} b \partial^2 \phi )~e^{\sqrt{3} \phi}
\eqno(2.22)
$$
For our purpose, the explicit forms are not so important. But as we have
seen, in general, the Lian-Zuckerman operators can be written as
$$
{\cal O}_{n,i,j}~=~w^n x^i y^j
\eqno(2.23)
$$
where $n\in Z$ and $-n$ refers to the ghost number of the operator. $i,j$ are
also integers with the restriction $0\leq i\leq p-2;~0\leq j\leq q-2$. This
restriction comes from the fact that the matter momentum of a physical
operator should lie inside the Kac table. Note that the Liouville momentum
does not necessarily remain inside the Kac table. The other fact to be noted is
that the operators $w$, $x$ and $y$ satisfy the commutation relations among
themselves as $w x = -x w$, $w y = - y w$ but $x y = y x$. Till now we
discussed only the chiral operators. In order to obtain the closed string
states we have to combine the holomorphic and anti-holomorphic operators in
such a way that their momenta match. Obviously, the solution is to take the
Lian-Zuckerman operators for $d < 2$ closed string theory as
$$
w^n x^i y^j {\bar w}^n {\bar x}^i {\bar y}^j
\eqno(2.24)
$$
Since the corresponding states are annihilated by both $b_0$ and ${\bar b}_0$
separately, these states belong to the so-called relative cohomology of the
closed string BRST operator. In the next section, we discuss the other physical
states for the closed string theory and the symmetry algebra associated with
them.

\noindent{\bf III. Symmetry Algebra of $d < 2$ Closed String States:}

It is well known in the operator formalism that the definition of a physical
state being annihilated by both $b_0$ and ${\bar b}_0$ is a stronger condition
than necessary [18]. On the contrary, what is necessary is that a physical
state
should be annihilated by $b_0 - {\bar b}_0$. As a consequence, besides the
operators in (2.24) which satisfy this condition trivially, there are more
states (or operators) which are also annihilated by $b_0 - {\bar b}_0$ and
hence are physical. This can be easily seen, as explained in [7], by the
existence of an operator
$$
a~=~[ Q, \phi ]~=~( c \partial \phi~+~i Q_L \partial c )
\eqno(3.1)
$$
and it's antiholomorphic counterpart
$$
{\bar a}~=~[ {\bar Q}, \phi ]~=~( {\bar c} {\bar \partial}\phi~+~i Q_L {\bar
\partial} {\bar c} )
\eqno(3.2)
$$
where ${\bar Q}$ is the antiholomorphic part of the BRST operator $Q$ as
defined in (2.12). Note that $a w~= - w a$ and $a$ commutes with $x$ and $y$.
Besides, $a$ and ${\bar a}$ have ghost number (1,0) and (0,1) respectively
and both of them have conformal weight zero. Thus, we have the new operators
$$
\Omega^{(0)}~=~(a + {\bar a})~w^n x^i y^j~{\bar w}^n {\bar x}^i {\bar y}^j
\eqno(3.3)
$$
which are also annihilated by $b_0 -{\bar b}_0$ and hence are physical
operators. Note that they are not independently annihilated by either $b_0$ or
${\bar b}_0$ i.e. they do not belong to the relative cohomology but belong to
the semi-relative cohomology of the full BRST charge. Following [8], one can
construct the conserved currents or the one-form $\Omega^{(1)}$ from the
operators in (3.3) by making use of the descent equation
$$
d \Omega^{(0)}~=~\{ Q +{\bar Q}, \Omega^{(1)} \}
\eqno(3.4)
$$
In other words, $\Omega^{(1)}$ is simply obtained upto BRST commutator by
multiplying (3.3) with $b_{-1}$ and ${\bar b}_{-1}$ separately. Since,
$\Omega^{(1)}~=~\Omega_z^{(1)} dz~+~\Omega_{\bar z}^{(1)} d{\bar z}$ where
$$
\eqalign{\Omega_z^{(1)}&=~{1\over 2\pi i}\oint_z dz' ~b (z') \Omega^{(0)} (z)
\cr \Omega_{\bar z}^{(1)}&=~{-1\over 2\pi i}\oint_{\bar z} d{\bar z}'~{\bar b}
({\bar z}') \Omega^{(0)} ({\bar z})\cr}
\eqno(3.5)
$$
the symmetry charges would be given by
$$
\Omega ~=~\oint \Omega^{(1)}
\eqno(3.6)
$$
In the above, we have chosen the convention that
$$
{1\over 2\pi i} \oint {d z\over z}~=~- {1\over 2\pi i} \oint {d {\bar z}\over
{\bar z}}~= 1
\eqno(3.7)
$$
Now, one can find the action of the currents or the charges on various
physical operators of the type (3.3) and we will give the explicit example
later. It is interesting to note that the conserved charge associated with
the physical operator $a + {\bar a}$ (i.e. operator of the type (3.3) with
$n = i = j = 0$) measures the Liouville momentum of a physical state. Since
the one-form associated with $(a + {\bar a})$ is
$$
\Omega^{(1)}~=~\partial \phi d z~+~{\bar \partial} \phi d {\bar z}
\eqno(3.8)
$$
so, the vector field of the corresponding conserved charge can be identified as
$$
 \sqrt{-2pq} ( w\partial_w~+~{1\over p} x\partial_x~+~{1\over q} y\partial_y
{}~-~{\bar w}\partial_{{\bar w}}~-~{1\over p} {\bar x}\partial_{{\bar x}}~-~
{1\over q} {\bar y}\partial_{{\bar y}} )
\eqno(3.9)
$$
It is clear that (3.9) will act trivially on all the physical states, since
they have matching left and right Liouville momenta. By working out a few
cases explicitly, it is easy to convince oneself that
the vector field associated with
the conserved charges of the operators $w^n x^i y^j$ is
$$
(-1)^n \sqrt{- 2pq} ( n + {i\over p} + {j\over q} ) w^n x^i y^j \partial_a
\eqno(3.10)
$$
Using (3.9) and (3.10), it is now straightforward to write down the vector
fields associated with the charges corresponding to both the operators of
the type (2.24) and (3.3). Throwing away some overall factors they are
respectively given by
$$
G_{n,i,j}~=~w^n x^i y^j {\bar w}^n {\bar x}^i {\bar y}^j (\partial_a~-~
\partial_{{\bar a}})
\eqno(3.11)
$$
and
$$\eqalign{K_{n,i,j}&=~w^n x^i y^j {\bar w}^n {\bar x}^i {\bar y}^j \left[ w
\partial_w + {1\over p} x \partial_x + {1\over q} y \partial_y\right.\cr &-~
( n + {i\over p} + {j\over q} ) a \partial_a + ( n + {i\over p} + {j\over
q} ) a \partial_{{\bar a}} - ( n + {i\over p} + {j\over q} ) {\bar a}\partial_a
\cr &\left.-~{\bar w}\partial_{{\bar w}} -
{1\over p} {\bar x} \partial_{{\bar x}} -
{1\over q} {\bar y} \partial_{{\bar y}} + ( n + {i\over p} + {j\over q} )
{\bar a} \partial_{{\bar a}} \right]\cr}
\eqno(3.12)
$$
It is now a simple exercise to compute the symmetry algebra of the conserved
charges and we find after introducing appropriate cocycle factors [9] that
$$
\eqalign{[ G_{n, i, j},~G_{m, k, l} ]&=~0\cr
[ G_{n, i, j},~K_{m, k, l} ]&=~0\cr
[ K_{n, i, j},~K_{m, k, l} ]&=~0\cr}
\eqno(3.13)
$$
Thus, the symmetry charges for the closed string physical states commute
among themselves. This is significantly different from the open string case
[9], where it is observed that the conserved charges $K_{n, i, j}$ satisfy a
Virasoro algebra when $i = j = 0$. Also, the algebra (3.13) does not agree
with the algebra conjectured in ref.[19]. The implication of the result (3.13)
is that, the currents corresponding to physical operators of the type (2.24)
and (3.3) annihilate the physical states. In general, it is very difficult to
have an explicit verification of this statement. But let us consider the (3,2)
model as a simpler example and see that this is indeed true. In this case, the
matter sector is trivial since the only matter field is the identity field.
Consider an operator of the type (3.3) with $i = j = 0$ and $n = - 1$ in this
model. Using the explicit form of $w^{-1}$ from $(2.21b)$ and $a, {\bar a}$
in (3.1), (3.2),
the physical operator $(a + {\bar a}) w^{-1} {\bar w}^{-1}$ takes the form
$$
(a + {\bar a}) w^{-1} {\bar w}^{-1}~=~{1\over 2\sqrt{3}} ( \partial c c {\bar
c}~+~{\bar \partial} {\bar c} c {\bar c} ) e^{- \sqrt{3} \phi}
\eqno(3.14)
$$
The conserved charge associated with this operator is found, using (3.5) and
(3.6), to be
$$
\Omega ~=~{1\over 2\pi i}\oint dz (\partial c {\bar c}~+ {\bar \partial} {\bar
c} {\bar c}) e^{- \sqrt{3} \phi}~-~{1\over 2 \pi i}\oint d{\bar z} ( \partial c
c~+ {\bar \partial} {\bar c} c ) e^{-\sqrt{3} \phi}
\eqno(3.15)
$$
where an overall factor of ${-1\over 2\sqrt{3}}$ is thrown out.
Working out the relavant OPEs, it is easy to check that $\Omega$ acts
trivially on $W = w {\bar w},~X = x {\bar x}$ and $(a + {\bar a})$. the
generator $Y = y {\bar y}$ is absent in this model. Thus, we conclude that
the symmetry charges of the $d<2$ free
closed string states act trivially on all the physical states.

\noindent{\bf IV. Conclusion}

The algebra of the symmetry charges (3.13) is not surprising in the sense that
one cannot costruct a non-trivial correlator with the free closed string
physical states ( or operators) given in (2.24) and (3.3). In order to have
a non vanishing correlator, one should be able to construct an operator with
ghost number six and matter and Liouville charge equal to the background
charges $\alpha_0$ and $\beta_0$. From the ghost number counting we note that
the above condition can be satisfied if one includes the operators of the type
$a {\bar a} w^n x^i y^j {\bar w}^n {\bar x}^i {\bar y}^j$. But, these operators
are not present in the closed string cohomology as they are not annihilited
by $(b_0 - {\bar b}_0)$.

To conclude we have shown here that the symmetry charges associated with
the discrete
states in $d<2$ free closed string theory
commute among themselves. This in turn implies that they act trivially on all
the physical states. It would be interesting to examine how this conclusion
is modified when the cosmological constant is taken to be non-zero and whether
the symmetry charges in that case reveal any interesting algebraic structure
in analogy with the matrix model results.
\vfil
\eject
\noindent{\bf Acknowledgments}

We thank M.H. Sarmadi for many valuable discussions. S.R. thanks Prof. A.
Salam,
    IAEA, UNESCO and ICTP, Trieste, for support. S.P.'s work is performed as
part of the research program of the ``Stichting voor Fundamenteel Onderzoek
der Materie'' (FOM).
{\titlea References}
\item{1.} B. Lian and G. Zuckerman, Phys. Lett. B254 (1991) 417; Comm. Math.
Phys. 135 (1991) 547.
\item{2.} F. David, Mod. Phys. Lett. A3 (1988) 1651.
\item{3.} J. Distler and H. Kawai, Nucl. Phys. B321 (1989) 509.
\item{4.} C. Imbimbo, S. Mahapatra and S. Mukhi, Nucl. Phys. B375 (1992) 399.
\item{5.} P. Bouwknegt, J. McCarthy and K. Pilch, Comm. Math. Phys. 145
(1992) 541.
\item{6.} D. Kutasov, E. Martinec and N. Seiberg, Phys. Lett. B276 (1992), 437.
\item{7.} E. Witten, Nucl. Phys. B373 (1992), 187.
\item{8.} E. Witten and B. Zwiebach, Nucl. Phys. B377 (1992) 187.
\item{9.} H. Kanno and M. H. Sarmadi, preprint IC/92/150 (1992).
\item{10.} V. Dotsenko, Mod. Phys. Lett. A6 (1991) 3601; Y. Kitazawa,
Phys. Lett. B265 (1991) 262.
\item{11.} S. Govindarajan, T. Jayaraman, V. John and P. Majumdar,
Mod. Phys. Lett. A7 (1992) 1063; S. Govindarajan, T. Jayaraman
and V. John, preprint IMSc-92/30.
\item{12.} D. Gross and I. Klebanov, Nucl. Phys. B344 (1990) 475; U. Danielsson
and D. Gross, Nucl. Phys. B366 (1991) 3.
\item{13.} M. A. Awada and S. J. Sin, Univ. of Florioda preprint IFT-HEP
-90-33; IFT-HEP-91-3.
\item{14.} S. R. Das, A. Dhar, G. Mandal and S. R. Wadia, preprint IASSNS-
HEP-91/52; Mod. Phys. Lett. A7 (1992) 71.
\item{15.} R. Dijkgraaf, G. Moore and R. Plesser, preprint IASSNS-HEP-92/8
YCTP-P22-92.
\item{16.} M. Bershadsky and I. Klebanov, Nucl. Phys. B360 (1991) 559.
\item{17.} S. Hosono, University of Tokyo preprint UT 595 (1992).
\item{18.} J. Distler and P. Nelson, Comm. Math. Phys. 138 (1991) 273.
\item{19.} S. Kachru, preprint PUTP-1314 (1992).

\bye